\documentclass{article}

\usepackage{arxiv}

\usepackage[utf8]{inputenc} % allow utf-8 input
\usepackage[T1]{fontenc}    % use 8-bit T1 fonts
\usepackage[hidelinks]{hyperref}       % hyperlinks
\usepackage{url}            % simple URL typesetting
\usepackage{booktabs}       % professional-quality tables
\usepackage{amsfonts}       % blackboard math symbols
\usepackage{nicefrac}       % compact symbols for 1/2, etc.
\usepackage{microtype}      % microtypography
\usepackage{cleveref}       % smart cross-referencing
\usepackage{lipsum}         % Can be removed after putting your text content
\usepackage{graphicx}
\usepackage[square,numbers]{natbib}
\usepackage{doi}
\usepackage{authblk}

\title{zkTax: A pragmatic way to support zero-knowledge tax disclosures}
\renewcommand{\shorttitle}{zkTax: zero-knowledge tax disclosures}

\author {
    \textbf{
    Alex Berke\textsuperscript{\rm * 1},
    Tobin South\textsuperscript{\rm * 1},
    Robert Mahari\textsuperscript{\rm 1 2},
    Kent Larson\textsuperscript{\rm 1},
    Alex `Sandy' Pentland\textsuperscript{\rm 1}
    }
    \\
   \textsuperscript{\rm 1}MIT Media Lab\\
   \textsuperscript{\rm 2}Harvard Law School\\
    \textsuperscript{\rm *} These authors contributed equally. \\
    aberke@mit.edu, tsouth@mit.edu
}

\renewcommand{\headeright}{}
\renewcommand{\undertitle}{}
\hypersetup{
pdftitle={zkTax: A pragmatic way to support zero-knowledge tax disclosures},
pdfauthor={Alex Berke, Tobin South, Robert Mahari, Kent Larson, Alex `Sandy' Pentland}
}

\begin{document}
\maketitle

\begin{abstract}
Tax returns contain key financial information of interest to third parties: public officials are asked to share financial data for transparency, companies seek to assess the financial status of business partners, and individuals need to prove their income to landlords or to receive benefits. 
Tax returns also contain sensitive data such that sharing them in their entirety undermines privacy. 
We introduce a zero-knowledge tax disclosure system (zkTax) that allows individuals and organizations to make provable claims about select information in their tax returns without revealing additional information, which can be independently verified by third parties.
The system consists of three distinct services that can be distributed: a tax authority provides tax documents signed with a public key; a Redact \& Prove Service enables users to produce a redacted version of the tax documents with a zero-knowledge proof attesting the provenance of the redacted data; a Verify Service enables anyone to verify the proof.
We implement a prototype with a user interface, compatible with U.S. tax forms, and demonstrate how this design could be implemented with minimal changes to existing tax infrastructure.
Our system is designed to be extensible to other contexts and jurisdictions. This work provides a practical example of how distributed tools leveraging cryptography can enhance existing government or financial infrastructures, providing immediate transparency alongside privacy without system overhauls.
\end{abstract}

% keywords can be removed

\section{Introduction}
% General Introduction
% Organizations and individuals frequently need to attest their financial information to third parties. 
% This can be accomplished by sharing tax returns, however, these documents also include sensitive private information. 
% Reliably and privately sharing sensitive data often poses a challenge in cooperative work contexts.
% Frequently in multi-party negotiations that involve financial dealmaking, verifying facts frequently becomes a key stumbling block. This issue typically prompts the need for costly insurance and legal protections. Such a pervasive problem in collaborative work underscores the necessity for provable assertions about private data.

Tax documents aggregate a wealth of information about an individual's finances, ranging from income sources to charitable contributions.
Upon submitting these documents to government agencies, individuals declare the information is true and accurate to the best of their knowledge. 
Similarly, organizations -- ranging from non-profits to corporations -- conduct financial audits and report their financial data through tax filings.
%Occasionally, individuals and organizations will need to share information that is part of their tax returns with third parties to secure loans, determine eligibility for public benefits, or to make financial disclosures.
% For example, in the US, individuals make such a declaration under the penalty of perjury upon submitting income tax returns to the IRS.

The information contained in tax returns, while primarily intended for tax-filing purposes, is often useful across a broad range of contexts. As a key motivating example, public officials can release tax details in the pursuit of increased accountability and transparency. 
Further, data derived from tax records can be instrumental in assessing eligibility for benefits, such as welfare benefits for low-income households~\cite{nichols2015earned, rosenbaum2013relationship}. Additionally, in legal proceedings such as divorce or child support negotiations, financial attestations are often substantiated through the use of this information~\cite{nichols2015earned, rosenbaum2013relationship}.

Yet, can these requests for financial records be satisfied in a verifiable way without exposing other private data present in tax records? Trustworthy mechanisms to verify the accuracy and authenticity of financial attestations are required.

% Problem statement
% Tax forms contain both financial information that is relevant to third parties and sensitive private data such as addresses, social security numbers, and detailed financial records.
% Can this data be released partially in a secure way without exposing private data?
% This requires trustworthy mechanisms to verify the accuracy and authenticity of released tax details.

%various fields that individuals might need to share for specific purposes, disclosing all data together presents privacy and information security risks. This is because tax forms include sensitive personal details, such as social security numbers (SSNs), and other information that may not be necessary to share and which individuals may prefer to keep private.
% TODO we should specify here what other information may be sensitive and why
%To securely release partial information, trustworthy mechanisms must be developed to verify the accuracy and authenticity of released tax details without exposing other information.

% Solution statement
We present a system to address this problem by leveraging zero-knowledge and public key cryptography. 
Zero-knowledge (ZK) in this context refers to the ability to computationally prove a statement is true, without exposing the inputs to the statement. 
% For example, ZK technologies can be used to prove that the public key signature on a tax return is valid or that an individual's income exceeds a certain amount without revealing the underlying data.
% We describe the system and then present a proof of concept prototype, publicly available at \url{http://zktax.media.mit.edu}. 
The system allows individuals to produce a redacted version of their tax documents such that the information can be verified by any third party. System users may either send the redacted documents to inquirers or publish the redacted versions for anyone else to publicly verify. Simple extensions of this system allow users to prove more complex statements about their tax returns.
This allows organizations and individuals to prove their financial information to third parties without revealing sensitive data. 

\subsection{Motivating Example}
Although we present multiple use cases for our system in \autoref{sec:discussion}, we start by highlighting a use case that will follow throughout demonstrating the system. 

This system is particularly motivated by the ability to leverage cryptographic tools to improve transparency and accountability in governance systems, of which the common practice of public officials releasing redacted versions of their income tax returns is central. 
For example, in the US, presidents have historically released their tax returns in a tradition of transparency. 
Proponents of measures to codify this act of transparency argue that public disclosure of tax returns could expose conflicts of interest, reveal annual tax liability and tax rates, and enable the public to observe whether the President or candidates have engaged in tax evasion, or pursued tax avoidance~\cite{prestaxreview}.
Yet, there have been recent cases where public officials have refused to release their taxes, despite demands from their own constituents and the general public~\cite{trumpnyt}. 
This work is motivated by the idea that if a zero-knowledge tax disclosure mechanism were available, this refusal to release requested information may not be necessary or defensible. In addition, the information disclosed could be better trusted through verification.

Consider a scenario where citizens inquire about just one field from a public official's tax forms, such as the total amount of tax paid. The official provides the requested amount, but refuses to release their tax returns due to privacy concerns. The citizens may then question whether the disclosed amount is truthful. The ZK tax disclosure system presented here offers a solution, allowing the official to provide verifiable proof of the specified value from their tax form without exposing sensitive information. This balance of privacy and transparency can potentially enhance public trust in governance systems through the application of Privacy by Design principles.

\section{Background and related work}
The system proposed here builds on both redactable signature schemes and recent advances in zero-knowledge proof tools. 
Here we elaborate on the foundational material of this work and contrast it to alternative systems that have been proposed.
 
% [History fo zkp]
Zero-knowledge proofs are a way to prove the validity of a statement without revealing any additional information beyond the truth of the statement itself~\cite{original}. 
Zero-Knowledge Succinct Non-Interactive Arguments of Knowledge (zk-SNARKs)~\cite{zksnark} have become a flexible ecosystem of proof generation software, with general-purpose high-level languages like circom~\cite{circom}, which compiles down to arithmetic equations called `circuits'. These circuits can then be used by powerful proving systems such as groth16~\cite{groth16} and halo2~\cite{halo} to produce small digitally verifiable proofs.

% [general related applications]
This work draws on these zk-SNARK advancements and their application to the financial system. These applications have focused on the blockchain space, such as their use in achieving anonymous transactions~\cite{zcash,TCwhitepaper}.
ZK proofs have also been used to achieve anonymous e-taxing protocols (using blockchains)~\cite{etaxing}, provable audits (zkLedger~\cite{zkledger}), and proof of reserves~\cite{provisions}. 
Other related work has also used blockchains to add privacy-preserving tools to tax systems, even without zero-knowledge proofs~\cite{weirdnorwaytaxpeople}

The works above assume a closed financial system where transactions occur on a blockchain. This is not today's reality. 
Unlike these prior works, our system does not assume a blockchain and can be implemented to work alongside and improve current tax and financial infrastructures.

%Zero-knowledge cryptography has been used in other applied computing contexts including ZK credential verification in negotiation~\cite{zkTrust}, digital identity assurance~\cite{CSCWdigitalIdentity}, and smart contract based distributed key generation~\cite{Sober2022DistributedKG}. 
ZK tools such as this work can play a powerful role in creating new user centric privacy pattern selections~\cite{AlMomani2021LandOT}. 
Indeed, since the introduction of GDPR there has been an increase in Privacy by Design research from a regulation-centric perspective~\cite{deChaves2023PrivacyBD}, but user-centric solutions still need research. While important work such as decentralized access control in blockchain-based systems~\cite{Jannes2023DEDACSDA} has already been done in this space, the work we present here provides a Privacy by Design approach where less technical users can be in control of their privacy tradeoffs while interacting with systems designed for the general public. 
 
% Similar themes of privacy, security, and portability have underpinned a long history of cooperative work techologies~\cite{CSCWconcepts} from early email design~\cite{CSCWemail} through to recent data sharing and custody approaches~\cite{CSCWblockchaindata,CSCWblockchainsecruity}. 

Other cryptographic approaches to providing verifiable redacted data have been taken, such as with content extractable signatures (CES)~\cite{steinfeld2002content} and redactable signature schemes (RSS)~\cite{sanders2020efficient}. 
In terms of our use case, we might consider a tax form as a set of messages, where each form field and its associated value is a message that can separately have an attached signature. A verifiable redaction scheme may then publish a subset of these signed fields. Works studying CES and RSS improve efficiency over providing a signature for each such message. Yet these approaches suffer from drawbacks where either the size of the signature or key must scale with the size of the original message (e.g. original tax document).
While this is not the case for our solution with zk-SNARKs, we note that the size of the circuit in our solution does grow with the allowed original tax document(s) size. However, unlike with CES and RSS, this burden does not then fall on the services that handle signing and is instead distributed.
Furthermore, unlike CES and RSS, our flexible approach with zk-SNARKs can be easily extended beyond simple redaction to prove relations between form field values while avoiding publishing the values themselves.

A more similar approach to this work is the use of zk-SNARKs to redact images by Ko et al~\cite{ko2021efficient}. They used a combination of zk-SNARKs to produce a ``public'' and ``private'' version of the image, and utilized the additive properties of a Pederson commitment scheme so that a verifier could verify that commitments on these two versions added up to a signed commitment from the original version. Given the complexity of their approach, signature verification was handled outside of a zk-SNARK, unlike our implementation. This process was designed specifically for images as represented by pixels, while our simpler approach can leverage the structured content of a tax form.

% They used a combination of zk-SNARKs to produce a "public" and "private" version of the image, and utilized the additive properties of a Pederson commitment scheme so that a verifier could verify that commitments on these two versions added up to a signed commitment from the original version. Given the complexity of their approach, signature verification was handled outside of a zk-SNARK, unlike our implementation. This process was designed specifically for images as represented by pixels, while our simpler approach can leverage the structured content of a tax form.

% [government public key infrastructure]
Our system assumes the existence of a digital service that uses public key cryptography to sign tax documents. We note this assumption is not far-fetched given the following work.
In 2005 Mexico took on an ambitious public key infrastructure project which allows citizens to sign tax declarations~\cite{mexicoPKI}, and Estonia's modern government infrastructure allows for a variety of digital signing including with taxes~\cite{eEstoniaNotAPaper}. In general, the global push towards Open Banking~\cite{openbanking} beckons a possible future where many financial providers will allow users to access permissions and signed attestations about financial products and balances. 
%Governments (especially in the western world) are slow to utilize new privacy technologies, often leaving it to the private sector or open source initiatives (e.g., g0v in Taiwan~\cite{govzero}) to provide innovation. This proposed system is an example of how modern cryptography and technologies can work to enhance government technology systems.

% s\subsection{Contributions}
% We describe a system that leverages ZK proofs to address important issues related to tax disclosures. This system applies a user-centric consentful privacy approach to the widespread problem of sharing tax information with third parties. The system is designed to integrate into existing tax systems without overhauling institutional structures.
% To demonstrate the feasibility of such a system, we provide an open-source implementation and prototype webapp. 
% The implementation includes ZK circuits that handle redaction and signature checking, proof generation, and proof verification, along with a user interface for accessibility to non-technical users. 

\section{zkTax System}

\begin{figure}[h]
    \includegraphics[width=0.95\linewidth]{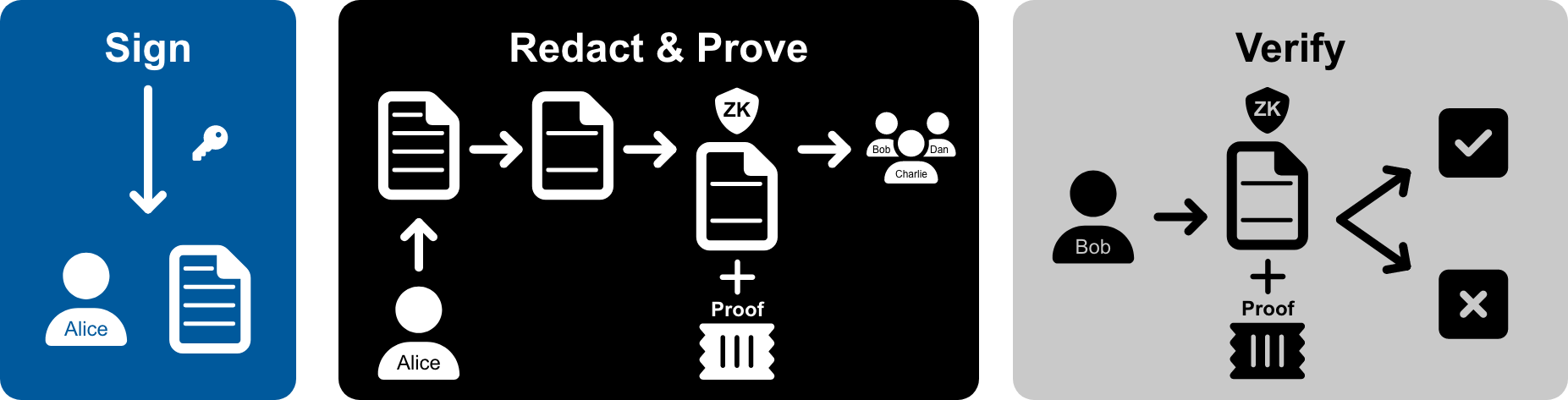}
    \caption{Diagram of the three services that compose the ZK tax disclosure system. (Left) An individual, Alice, retrieves tax data from a Trusted Tax Service (TTS), which signs the tax data with a signature that can be verified with a public key. (Middle) Alice brings the signed tax data to the Redact \& Prove Service. The service allows her to select how tax data should be redacted and then generates a redacted version and proof that the original data was signed by the TTS. Alice can make the proof and redacted data public for anyone to then verify by using the Verify Service.
    \label{fig:diagram_3services}}
\end{figure}

% We take a pragmatic approach to designing a system that could be readily implemented by tax authorities and used by individuals and organizations. Furthermore, the system brings privacy choices directly to the user for consentful private sharing of data without technical expertise.

\subsection{System goals}
To enable and enhance usability the system should achieve the following.
\begin{itemize}
  \item \emph{Interoperability with existing tax services}: 
  Building on top of existing taxation infrastructure minimizes the burden for tax administrators while facilitating adoption.
  % Seamless integration with current tax systems to ensure a smooth transition and adoption. 
  % This integration should require as few interactive steps as possible on behalf of the tax service provider (e.g. tax data should be signed once). 
  \item \emph{Non-interactive from the perspective of the tax service}: 
  Tax service providers (e.g. government agencies) should sign tax data once to enable users to generate verifiable attestations from this data; minimizing costs for service providers and mitigating likely bottlenecks.
  % This minimizes costs and inefficiencies for tax service providers.
  % The tax service provider should be able to sign the tax data only once before handing off the protocol, in order to minimize the burden and and opportunity for a bottleneck within the tax service provider (e.g. government agency).
  \item \emph{Public verifiability}: Third parties should be able to independently verify the redacted tax information. This should be user-friendly for end-users without a background in cryptography.
  \item \emph{Flexibility and extensibility in terms of what can be proved}: The system should be adaptable to accommodate different user desires for combinations of attestations and future extensions.
  \item \emph{Modularity}: A modular design simplifies the system's maintenance, upgrades, and potential integration with other systems.
  \item \emph{Privacy by design for the taxpayer}: The system should not reveal unauthorized private data to any third party except the trusted tax service and protect against re-identification attacks. 
  % No disclosure of private tax data to anyone or any service aside from the trusted tax service, beyond what the taxpayer has chosen to disclose. This includes reverse-engineering proof artifacts to infer full or partial tax inputs.
  \item \emph{Unforgeability}: Individuals should be unable to fraudulently claim that redacted tax information was sourced from real tax information handled by a tax service.
  % At no point should sensitive data need to be released to the public or uploaded to another service. This also means that proof artifacts must not be able to be reverse-engineered to infer full or partial tax inputs.
  % We note that another solution that achieves verifiable redaction might be to ask a government tax agency (e.g. the IRS in the US) to supply the redacted version. However, this is not a common practice and would place undue burden on the tax system, and could introduce bureaucratic friction that would impede the goals of transparency and accountability. The system we present is easily and quickly usable once a taxpayer has received their signed tax data. 

  % \item \emph{Self-sovereignty of private data}: Empowers individuals to control the disclosure of their sensitive information. Disclosure controls are not one-size-fits-all.
\end{itemize}

\subsection{Architecture}

To achieve the above goals, we propose a system consisting of three distinct services. 
A high-level view of this architecture is illustrated in Figure~\ref{fig:diagram_3services}. Figure~\ref{fig:systemoverview} provides further detail of the system components and how data flows between them.
This section describes the proposed architecture at a high level, where the specifics of implementation can vary. 
Section~\ref{sec:demonstration} describes our prototype implementation in further detail.

\begin{table*}
\small
\begin{tabular}{ | p{0.15cm} | p{1.5cm} | p{6cm} | p{1.5cm} | p{6cm} |}
\hline
& \multicolumn{2}{|l|}{System participants} & \multicolumn{2}{|l|}{Services}        \\
\hline
& Name & Description & Name & Description \\
\hline
1 & Trusted tax authority & Entity that already has access to taxpayers' tax data. E.g. in the U.S. this may be the IRS, or a company that processes taxes and sends them to the IRS on behalf of taxpayers (e.g. TurboTax). & Trusted Tax Service (TTS) & Returns tax data to individuals, where the data is signed and can be verified using a public key published by the TTS. \\
\hline
2 & Individual taxpayer & Individual who retrieves their signed tax data from the TTS and uses the Redact \& Prove Service to redact fields from the data and produce a ZK proof about the redacted data. The full tax data is private to them. They share the redacted data. & Redact \& Prove \newline Service & Shares a zk-SNARK circuit with the Verify Service. Provides an interface for taxpayers to privately select fields to redact from their tax data and produces a proof of the redacted data by using the zk-SNARK. \\
\hline
3 & General \newline public or \newline auditor & Any entity who wishes to verify a claim about the redacted tax data. They take the shared redacted tax data and proof to a Verify service. & Verify \newline Service & Shares a zk-SNARK circuit with the Redact \& Prove Service. Provides interface for anyone to take shared redacted tax data and corresponding proof, output by the Redact \& Prove Service, to verify. \\
\hline
\end{tabular}
\caption{Overview of the system participants and their corresponding services in the proposed tax verification system.\label{tab:system_roles_summary}}
\end{table*}

Table~\ref{tab:system_roles_summary} summarizes the distinct participant roles that correspond to the three services. 
% These are further described below.

% System participants:
% \begin{itemize}
% \item  Trusted tax authority: They are responsible for receiving and processing tax submissions from individual taxpayers. In our system, we refer to them as the ``Trusted Tax Service" (TTS). They have the ability to transform the tax data they process into a machine-readable format, sign it to produce a signature that is verifiable using a known public key, and return the signed, machine-readable tax data to the taxpayer. This allows other entities to verify that the tax data was approved by them.

% \item Taxpayer: The taxpayer is an individual who is responsible for submitting their tax information. This could be any person who wishes to state specific facts about their tax return. The taxpayer interacts with the TTS. The TTS transforms their tax data into a machine-readable format, signs it, and returns it to the taxpayer for further use. This taxpayer can use the ``Redact \& Prove Service" to redact sensitive information from their tax data and generate a proof that the redacted data is valid and originated from the complete tax data provided by the TTS.

% \item General public or another third party: This participant represents any individual or entity that wishes to verify a claim about a taxpayer's records. This could be a member of the public or an auditor automatically reviewing tax record proofs. They interact with the ``Verification Service", which allows them to confirm that the redacted taxes provided by the taxpayer did indeed originate from tax data signed by the TTS.
% \end{itemize}

In short, zk-SNARK circuit used to generate and verify proofs is publicly shared among the Redact \& Prove and Verify Services.
A taxpayer receives their signed and machine-readable tax data from the Trusted Tax Service (TTS), and then uses the Redact \& Prove Service to create a redacted version of their tax data along with a proof of its validity. This redacted data and proof can then be verified by any third-party using the Verify Service. 

\subsubsection{Trusted Tax Service (TTS)}
Our system assumes that there is a trusted authority -- which we refer to as the Trusted Tax Service (TTS) -- to which individuals submit their taxes. Such an entity might be the government or a trusted tax preparer that submits taxes to the government on an individual's behalf. 
Possible examples in the American tax system include the IRS, TurboTax, and H\&R Block. 
Our proposal assumes these services have the ability to transform the tax data they process into a machine-readable format (e.g. JSON), have a known public key, and can return the machine-readable version of the accepted tax data, along with a signature and their public key, so that other entities can verify the tax data was approved by them. 
We note the TTS assumed in this work is not currently implemented, yet is reasonable in scope. The IRS published a multiyear modernization plan in 2019 that described a taxpayer payment Application Programming Interface (API) to support WebApps~\cite{IRSmodernizationPlan}. Intuit currently provides an API~\cite{intuitAPI} for third-party developers to build apps using QuickBooks data in order to publish apps to the QuickBooks app store.

While ZK proofs may not be in the scope of software that these trusted entities will provide in the immediate future, our proposed system allows others to use the signed data returned from the TTS to build ZK proof systems that enable the privacy-preserving disclosures described. Although this approach does rely on existing entities returning signed tax data, we believe this kind of technical progress to be more possible in the short run than alternative solutions which have suggested complete replacement of existing tax services (such as with a distributed ledger~\cite{Niu2022ABC}).

In our proposed system, we assume the TTS provides a taxpayer's complete and unredacted tax information as JSON ($x$). 
The TTS uses the hash-then-sign paradigm by using a hash function ($H$) to produce a hash of the user’s complete tax information $H(x) = y$ and then signs $y$ using the TTS secret key ($sk$) as $S = sign(sk, y)$. 
The TTS returns $x$ and $S$ to the user. 
Note this data is private to the taxpayer, as it still contains the unredacted tax information, $x$.
$H$ and the TTS public key, $pk$, are publicly known.  

\subsubsection{Redact \& Prove Service}
A taxpayer takes the tax data ($x$) and corresponding signature ($S$) output from the TTS to a Redact \& Prove Service as inputs. This service may be run by any entity, but does require a certain level of trust because the input to the front end of this service is private tax data ($x$). This service can and should work by operating locally within the user's machine, to avoid sending private data to a server. This can be done by compiling all assets (including files to make proofs from the zk-SNARK circuit) beforehand and serving these files to the user's browser.

\begin{figure*}
    \includegraphics[width=\textwidth]{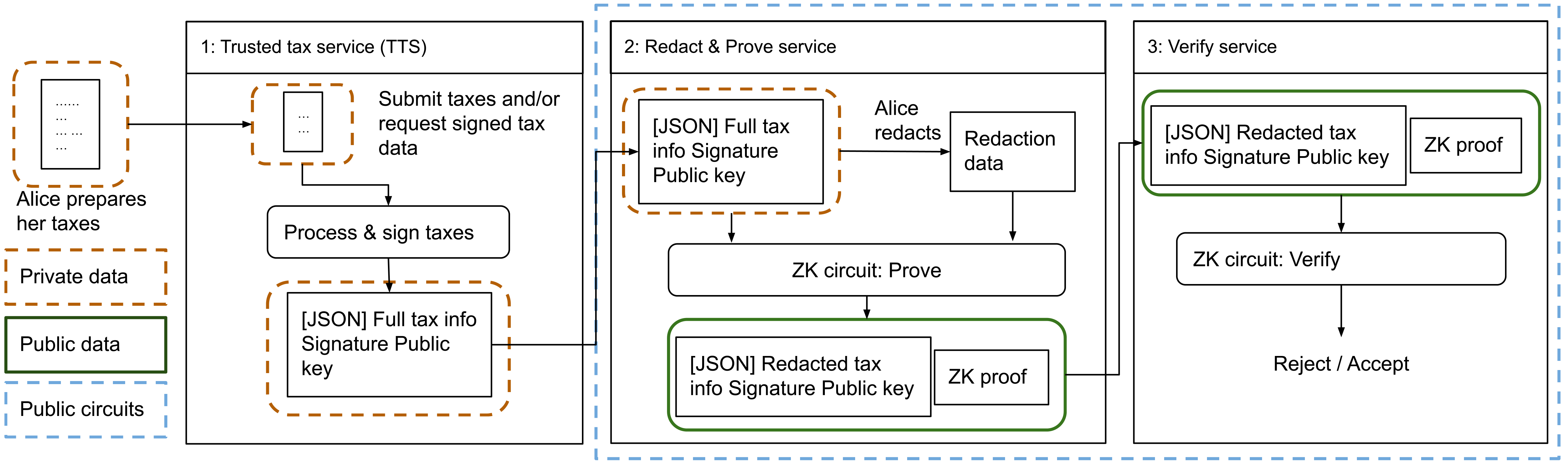}
    \caption{ZK Tax system overview with data flows.}
    \label{fig:systemoverview}
\end{figure*}

The Redact \& Prove Service operates in 2 steps. An interface allows the user to insert their tax data from the TTS and select where redactions should be made, producing an intermediate output: redact data. The proof generation step then takes 2 inputs: The output ($x$, $S$) from the TTS and the redact data. 
These inputs are provided to a zk-SNARK circuit that does the following. The circuit first verifies that the input tax data, $x$, matches the signature, $S$, by checking the hashed tax data against the TTS signature and public key: $verify(H(x), pk, S)$. If this verification fails, an error is returned to indicate that the tax information $x$ and signature $S$ did not match.
Otherwise, if the tax data $x$ are verified as valid, the circuit redacts all data indicated by the redact data input from $x$ to produce $x'$. The circuit output is both the redacted version of the data, $x'$, and a proof, $\pi_{x'}$. The Redact \& Prove Service returns $\pi_{x'}$, $x'$, $S$, and $pk$ so that others can check the validity of $x'$ (shown as the Reject/Accept output in Figure \ref{fig:systemoverview}). The original unredacted tax data, $x$, is the witness for the proof.

\subsubsection{Verification service}
The verification service allows for anyone to verify that the redacted tax data, $x'$, did indeed originate from some unredacted tax data, $x$, that was signed by the TTS with with public key $pk$. It does so by compiling the public circuit and verifying the compiled circuit, proof data ($\pi_x'$), and redacted data ($x'$) are compatible. Since the circuit code includes verifying the signature with $pk$ on the full tax data, $x$, this also means that the redacted tax data must match what the TTS returned.

% Barring the limitation that this system must integrate with existing tax services that are not already signing documents (discussed in \autoref{sec:disscussion}), this system addresses all of the system goals outlined above.

\section{Proof of concept implementation with U.S. tax forms}\label{sec:demonstration}

We developed a proof of concept implementation which includes a public web app\footnote{
\url{http://zktax.media.mit.edu}
% \url{https://18.27.79.206/} (converted to IP address for blind submission, certificate error may occur)
}, a set of reference circuits, and tax form processing code.
All code for the prototype is open source\footnote{
  % \url{http://REDACTED}
\url{http://github.com/tobinsouth/zktax}
}.

The web app is implemented using React. We implement the ZK proof and verification by leveraging the Circom software ecosystem~\cite{circom} to write and compile R1CS circuits. We use the snarkjs javascript library~\cite{snarkjsgithub} to handle the trusted setup, to produce the circuit artifacts used for proof and verification, as well as for implementing proof and verification within the user's browser. 

% In this section we first address the motivating use case.
% We then explain how data are represented in order to flow through the 3 services (TTS, Redact \& Prove, Verify) and the circuit.
% We then describe our zk-SNARK R1CS circuit and walk through the example use case with our prototype.

\subsection{Example Use Case}\label{sec:exampleusecase}

To demonstrate how this system can be used, we return to the motivating use case where a public official wishes to make a provable statement about specific information from their tax returns, without sharing other information.
In this example we use real data from a public official's released tax forms\footnote{Original forms available in the GitHub repository.
% \url{https://REDACTED}
% \url{https://github.com/tobinsouth/zktax/blob/main/app/public/f1040/f1040-2020-trump-downloaded-non-machine-readable.pdf}
}.
We assume the official wishes to redact all information except ``Taxable income" (field 15 on Form 1040) and their name. The resulting redacted form is shown in the Appendix Figure~\ref{fig:trumpf1040_redacted}.

% We consider a case where the public has asked a public official to share their ``Taxable income" (field 15 on the form).
% In this case, all fields are then redacted from the form, excluding the requested field (field 15) and the name fields, which are kept in order to associate the official's identity with the data. Our implementation handles redaction by setting all redacted fields to whitespace. 
% The resulting redacted Form 1040 is shown in the Appendix Figure~\ref{fig:trumpf1040_redacted}.

\subsection{Data representation}
Tax forms are often PDFs. We convert PDFs to and from a flat JSON representation, where form fields map to values, with metadata for form name and year assumed to be added by the TTS. The JSONs are converted to ASCII strings as input for the zk-SNARK circuit.

{\tt \small
\begin{verbatim}
json: {
// all data from the tax form
…
"fname": "name",
	"lname": "name",
"SSN": "000000000",
…
"f_1": "393,229", // represents field 1
"f_2a": "2,208", // represents field 2a
…
// additional metadata
"year": "2020",
"form": "1040"
}
\end{verbatim}
}

% In order to sign the data as well as provide the data as input to the zk-SNARK circuit, the JSON is stringified and converted into an array of ASCII characters.
% The data can be converted between the JSON and ASCII array representations, and therefore between the ASCII array and PDF representations as well.
In our interface, these machine data representations are converted into human-readable JSONs and filled PDFs.

Our circuit uses an array of bits to represent the redaction data produced by the taxpayer using the Redact \& Prove Service, where each element in the array indicates whether the corresponding element in the stringified JSON ASCII array should be redacted or not.

% The complete input to the circuit is represented below as JSON.
% {\tt \small
% \begin{verbatim}
% {
% 	"json": [ ASCII array representing JSON tax form data],
% 	"redact_map": [ bit array indicating elements to redact ],
% 	// TTS public key
% 	"pubkey": [ x, y ],
% 	// signature components
% 	"signature_R8x": value,
% 	"signature_R8y": value,
% "signature_S": value
% }
% \end{verbatim}
% }

% The output of the circuit is also a stringified JSON ASCII array representing the redacted json tax data.

\subsection{zk-SNARK circuit}

The zk-SNARK is implemented as an R1CS circuit with the following simple logic: the JSON input is hashed and the circuit utilizes the TTS public key (referred to as "pubkey") to assert that the input signature components correspond with the hashed JSON. If the signature verification is unsuccessful, a proof is not generated. Given a valid signature, the circuit yields an output representing the redacted JSON. This output aligns with the input JSON except where indicated by the redaction map bit array, which points to the corresponding elements set to whitespace.

We use EdDSA/MiMC-7 for computing hashes and signatures~\cite{eddsamimc}.
The MiMC hash function is chosen for its efficiency within zk-SNARK circuits, which is due to its low multiplicative complexity~\cite{mimc}.
We note consistent hash and sign functions must be used within the circuit and TTS.

Groth16 is used as the proving system~\cite{groth16} requiring a circuit trusted setup which is handled by using values computed from the Perpetual Powers of Tau ceremony~\cite{announcingpowerstau}.

Once run, the Redact \& Prove setup produces a computational \texttt{proof.json} and a \texttt{signals.json}, the latter of which contains the redacted JSON string as ASCII and the signer public key.

% Code for our circuit is shown in the Appendix Figure~\ref{fig:circuit}. The code is written and compiled using Circom.
% snarkjs~\cite{snarkjsgithub} is used to compile the circuits to JavaScript artifacts and then generate and verify the proofs. 

\subsubsection{Circuit size complexity}
The circuit is parameterized by the maximum JSON string input size such that the size of the circuit grows with the size of the input. For larger circuit sizes, the compilation needs larger Powers of Tau, and produces larger verification files. 
Our prototype uses a maximum JSON size of 1500.
Our open source code provides files for other input sizes.

\subsection{Prototype web app}
Our initial implementation is based on a US income tax form: the 2020 IRS Form 1040 provided \footnote{\url{https://www.irs.gov/pub/irs-prior/f1040--2020.pdf}}. The web interface for each service is shown in the appendix.
Our implementation includes a proof-of-concept TTS to which a user may upload a completed Form 1040 which is parsed into JSON, stringified, and hashed using MiMC before an EdDSA signature is generated from \texttt{circomlibjs}\footnote{\url{https://github.com/iden3/circomlib/blob/master/circuits/eddsa.circom}} which is returned, along with other necessary outputs, for the Redact \& Prove Service.
% Our prototype provides a link to the blank version of this form, so that others can try it out by filling in information. We also provide examples of filled out forms, based on those disclosed by public officials\footnote{\url{https://github.com/tobinsouth/zktax/blob/main/app/public/f1040/f1040-2020-trump.pdf}}.

% \subsubsection{Prototype TTS}

%We implemented a mock TTS to stand in the place of a TTS for the purpose of demonstration.
%A user can insert a filled out Form 1040. This is parsed into JSON, as shown side-by-side with the PDF.
% The service also transforms the JSON into the stringified ASCII array. This array must match the string size of 1500 characters expected by the circuit. The array is padded with the whitespace character to reach this size.
%Upon clicking a button, the service hashes the stringified JSON using MiMC, generates the EdDSA signature from \texttt{circomlibjs}\footnote{\url{https://github.com/iden3/circomlib/blob/master/circuits/eddsa.circom}} and returns this with other outputs necessary for the Redact \& Prove Service.

% \subsubsection{Prototype Redact \& Prove Service}

The Redact \& Prove Service interface generates the redaction bit array based on the user's selection of fields and the TTS output.
Note that this service is privately used by the taxpayer, given the input information contains all of their private tax data.
The zk-SNARK circuit files, which are used for redaction, are precompiled using snarkjs and served to the frontend. 
snarkjs is also used for the groth16 proof computation. 
All operations are done within the client browser.
The user can download these proof artifacts and share them as they contain no private data.

% Figure \ref{fig:redact_prove} shows the prototype Redact \& Prove Service interface. 
%Using the Redact \& Prove Service interface, users take output from the TTS to this service, which parses the stringified JSON to show the PDF and provides an interface to select fields for redaction.
%From the selected redaction fields, the service creates the redaction bit array and provides this, along with the data from the TTS, as input for the zk-SNARK circuit.
%Note that this service is privately used by the taxpayer, given the input information contains all of their private tax data.
%The zk-SNARK circuit files used are precompiled using snarkjs and served to the frontend. snarkjs is also used for the groth16 proof computation. All operations are done within the client browser.
% The outputs from the proof are the \texttt{proof.json} and \texttt{signals.json}. The signals contain the redacted version of the stringified JSON, produced by the circuit as described above.
%The user can download these proof artifacts and share them as they contain no private data.

% \subsubsection{Prototype Verify Service}

Similarly, the prototype Verify Service precompiles a verification key from the circuit via snarkjs using groth16 that is served to the service's frontend. Users can provide the \texttt{proof.json} and \texttt{signals.json} output from the proof. 
The verification key is then used to verify that the signals and proof are compatible, thereby accepting or otherwise rejecting the proof.

%In this Verify Service service, a verification key is precompiled from the circuit via snarkjs using groth16 and is served to this service's frontend. 

\section{Discussion}\label{sec:discussion}

\subsection{Extensible circuit logics}
This prototype focuses on using zero-knowledge proofs to prove that redacted tax data originated from official signed tax forms. 
While this functionality has numerous applications, extensions to this prototype, such as range-proofs and various logical assertions, could also be encoded into R1CS circuits.
Zero Knowledge Range Proofs (ZKRP), such as proposed by Boudot~\cite{boudot2001}, could allow a user to assert that a value, such as total taxes paid, falls within a given range.
Meanwhile, circuits can be designed to support a wide array of logical assertions, including but not limited to, statements like `the value of field A is greater than that of field B', or `the charitable donations made exceed a certain percentage of the income earned'. 
Services 2 and 3 have the capacity to incorporate ZK circuits that are tailored for each of these specific scenarios.
In line with our system goals, our implementation of a redaction service and circuit is aimed at generalizability and does not require specific fields to be hard-coded. 
The resulting flexibility enhances our ability to adapt to different use-cases without necessitating substantial alterations to the underlying circuitry.

\subsection{Example Use Cases}\label{sec:exampleuse}
Extending beyond the general prototype example and its application to a public official's disclosure, we highlight two additional contexts where zkTax allows individuals or organizations to prove financial information from their tax returns to third parties without sharing sensitive data. 

\subsubsection{Efficient and verifiable individual tax data attestations}

Individuals may need to prove their income or other financial information to potential landlords, employers, and lenders. To this end, they can rely on tax authorities (e.g. IRS) as a TTS and third-party providers to redact and prove their tax return information. The individual can freely share authenticated financial data and associated proofs with any requester (e.g. landlord or bank), who can confirm the authenticity using any third-party verification service.

In this way, zkTax approaches provide a trustworthy and privacy preserving way to share financial data, and when combined with range proofs, obviating the need for exact numbers to be shared directly, instead proving that a given financial figure is above or below a specific threshold. 

In addition to individuals and corporations attesting their financial information, this system could streamline the process of eligibility checking for government benefits. Governmental systems often separate data to maintain privacy, meaning that a welfare or benefits agency may not have access to original tax records. For welfare recipients who file tax returns, the proposed tool could make it easier to prove eligibility for benefits.

Not only does this approach minimize the amount of private information that is needlessly transmitted, it also reduces the cost of verification for the recipient, and eliminates the risk of human errors during verification.

\subsubsection{Verifying business counterparty risk}

Business transactions often involve counterparty risk —- the risk that the other party in a deal will be unable to fulfill its contractual obligations. Assessing counterparty risk generally involves reviewing financial data, but this may expose proprietary business information to competitors. 

A business can rely on their tax auditors (e.g., EY or KPMG) to serve as its TTS. When they enter into a new transaction, the counterparty can ask the business to share certain financial items from its tax return. In many cases, it will be sufficient for the business to provide range proofs or to prove relationships between different financial items (e.g. liquidity ratio or accounts payable days). Using the zkTax approach, a business can assemble the requested information and associated proofs to put its transaction partner at ease while minimizing unnecessary disclosures that could impact its competitiveness.

\subsection{Implementation requires minimal system overhaul and can be distributed}
Our implementation is a single web app to provide all three services,  Only (1) the trusted tax service (TTS) that accepts and signs the taxes needs to be operated by a specific trusted tax authority (that is already entrusted with sensitive tax information).
The (2) Redact \& Prove and (3) Verify Services can be operated by one or several other entities. 
This flexibility is a primary advantage of the modular system architecture we propose.
We believe it is a reasonable expectation for a trusted tax service, such as the IRS, to return signed JSON representations of individuals' accepted tax data to those individuals. 
Yet whether or not they also choose to provide zero-knowledge proof and verify services, such as those described in this work, does not matter for the system's feasibility. Any other entity can then implement services (2) and (3).

% We note that much of the related and prior work in auditing financial transactions assumes that transactions occur within a closed system, such as on a blockchain. 
% While these prior proposed systems can then offer many benefits and can avoid relying on a trusted third party, such as the IRS, they would require a complete system overhaul.
% Our proposal does not impose this requirement. As long as the trusted tax service returns signed data, implementation is feasible.

\subsection{Limitations}
A limitation of the proposed system is its reliance on a Trusted Tax Service (TTS) to sign tax documents. Many tax authorities currently do not have mechanisms for digitally signing individual tax documents in a manner compatible with our system. While the architecture is designed to minimally impact existing tax infrastructure, this requirement may pose an adoption hurdle. It is conceivable that interim solutions could be deployed, such as using email signatures from tax authorities~\cite{gupta2022zk} as a form of less secure but more readily available authentication.

Second, this work does not incorporate a user study involving real tax data to understand individual expectations and demands for privacy. Such a study would be essential for capturing the nuances of user interaction with the system and evaluating its practical usability. Moreover, the system's acceptance also hinges on the willingness of third-party requesters to rely on zero-knowledge proofs for data verification. This aspect remains an unexplored variable that could significantly influence the system's possibility of adoption.

% \subsection{Broader tax systems}

% Proving that the math has been done correctly.

% \subsection{Extentions to a coporate environment}
% Beyond the context of claims about tax paid, coporate 
% big 4 could allow companies to redact and prove their financial records for sharing in deal making

% \subsection{Goverment don't need to audit any more}
% Roberts crazy idea

\section{Conclusion}
We present a system that leverages zero-knowledge proofs and public key cryptography to address the challenge of verifying tax information without exposing sensitive data. Our system comprises three distinct services: a Trusted Tax Service (TTS), a Redact \& Prove Service, and a Verification Service. Together, these services enable taxpayers to make provable claims about their tax data that can then be verified by third parties. The architecture requires minimal changes to existing tax systems, offers flexibility in its deployment, and can be extended to support a variety of claims about financial data.

We demonstrate the feasibility of our system through a publicly available proof-of-concept implementation. This work contributes to financial privacy and data sovereignty for individuals, enhancing political transparency and accountability through public officials' financial disclosures, and allows businesses to assess counterparty risk without requiring proprietary or sensitive information.

%and governance by offering a practical solution to the challenge of verifying tax information in a privacy-preserving manner.


\section*{Availability}

The prototype with a public web interface is available at 
\url{http://zktax.media.mit.edu}
% \url{https://18.27.79.206/}
% \footnote{Converted to IP address for blind submission; certificate error may occur.}
All code for the prototype web app, zk-SNARK circuits, and tax form processing code, is available at
% \url{http://REDACTED}.
\url{http://github.com/tobinsouth/zktax}.

\section*{Acknowledgments}
We thank 
% [REDACTED] 
Husdon Hooper 
for his work on the visual design of the prototype web application.

\bibliographystyle{ACM-Reference-Format}
\bibliography{references} 


\begin{thebibliography}{33}

%%% the Web address.

\ifx \showCODEN    \undefined \def \showCODEN     #1{\unskip}     \fi
\ifx \showDOI      \undefined \def \showDOI       #1{#1}\fi
\ifx \showISBNx    \undefined \def \showISBNx     #1{\unskip}     \fi
\ifx \showISBNxiii \undefined \def \showISBNxiii  #1{\unskip}     \fi
\ifx \showISSN     \undefined \def \showISSN      #1{\unskip}     \fi
\ifx \showLCCN     \undefined \def \showLCCN      #1{\unskip}     \fi
\ifx \shownote     \undefined \def \shownote      #1{#1}          \fi
\ifx \showarticletitle \undefined \def \showarticletitle #1{#1}   \fi
\ifx \showURL      \undefined \def \showURL       {\relax}        \fi
% invisible to TeX
\providecommand\bibfield[2]{#2}
\providecommand\bibinfo[2]{#2}
\providecommand\natexlab[1]{#1}
\providecommand\showeprint[2][]{arXiv:#2}

\bibitem[Al-Momani et~al\mbox{.}(2021)]%
        {AlMomani2021LandOT}
\bibfield{author}{\bibinfo{person}{Ala'a Al-Momani}, \bibinfo{person}{Kim Wuyts}, \bibinfo{person}{Laurens Sion}, \bibinfo{person}{Frank Kargl}, \bibinfo{person}{Wouter Joosen}, \bibinfo{person}{Benjamin Erb}, {and} \bibinfo{person}{Christoph B{\"o}sch}.} \bibinfo{year}{2021}\natexlab{}.
\newblock \showarticletitle{Land of the lost: privacy patterns' forgotten properties: enhancing selection-support for privacy patterns}.
\newblock \bibinfo{journal}{\emph{Proceedings of the 36th Annual ACM Symposium on Applied Computing}} (\bibinfo{year}{2021}).
\newblock


\bibitem[Albrecht et~al\mbox{.}(2016)]%
        {mimc}
\bibfield{author}{\bibinfo{person}{Martin Albrecht}, \bibinfo{person}{Lorenzo Grassi}, \bibinfo{person}{Christian Rechberger}, \bibinfo{person}{Arnab Roy}, {and} \bibinfo{person}{Tyge Tiessen}.} \bibinfo{year}{2016}\natexlab{}.
\newblock \showarticletitle{MiMC: Efficient encryption and cryptographic hashing with minimal multiplicative complexity}. In \bibinfo{booktitle}{\emph{Advances in Cryptology--ASIACRYPT 2016: 22nd International Conference on the Theory and Application of Cryptology and Information Security, Hanoi, Vietnam, December 4-8, 2016, Proceedings, Part I}}. Springer, \bibinfo{pages}{191--219}.
\newblock


\bibitem[Baylina and Bell{\'e}s({[n.\,d.]})]%
        {eddsamimc}
\bibfield{author}{\bibinfo{person}{Jordi Baylina} {and} \bibinfo{person}{Marta Bell{\'e}s}.} \bibinfo{year}{[n.\,d.]}\natexlab{}.
\newblock \bibinfo{title}{EdDSA For Baby Jubjub Elliptic Curve with MiMC-7 Hash}.
\newblock
\newblock


\bibitem[Bell{\'e}s-Mu{\~n}oz et~al\mbox{.}(2022)]%
        {circom}
\bibfield{author}{\bibinfo{person}{Marta Bell{\'e}s-Mu{\~n}oz}, \bibinfo{person}{Miguel Isabel}, \bibinfo{person}{Jos{\'e}~Luis Mu{\~n}oz-Tapia}, \bibinfo{person}{Albert Rubio}, {and} \bibinfo{person}{Jordi Baylina}.} \bibinfo{year}{2022}\natexlab{}.
\newblock \showarticletitle{Circom: A Circuit Description Language for Building Zero-knowledge Applications}.
\newblock \bibinfo{journal}{\emph{IEEE Transactions on Dependable and Secure Computing}} (\bibinfo{year}{2022}).
\newblock


\bibitem[Ben-Sasson et~al\mbox{.}(2014)]%
        {zcash}
\bibfield{author}{\bibinfo{person}{Eli Ben-Sasson}, \bibinfo{person}{Alessandro Chiesa}, \bibinfo{person}{Christina Garman}, \bibinfo{person}{Matthew Green}, \bibinfo{person}{Ian Miers}, \bibinfo{person}{Eran Tromer}, {and} \bibinfo{person}{Madars Virza}.} \bibinfo{year}{2014}\natexlab{}.
\newblock \showarticletitle{Zerocash: Decentralized Anonymous Payments from Bitcoin}.
\newblock \bibinfo{journal}{\emph{2014 IEEE Symposium on Security and Privacy}} (\bibinfo{year}{2014}).
\newblock


\bibitem[Blank(2021)]%
        {prestaxreview}
\bibfield{author}{\bibinfo{person}{Joshua~D. Blank}.} \bibinfo{year}{2021}\natexlab{}.
\newblock \bibinfo{title}{Presidential Tax Transparency}.
\newblock \bibinfo{howpublished}{\url{https://yalelawandpolicy.org/presidential-tax-transparency}}.
\newblock
\newblock
\shownote{Accessed: 5-16-2023}.


\bibitem[Boudot(2000)]%
        {boudot2001}
\bibfield{author}{\bibinfo{person}{Fabrice Boudot}.} \bibinfo{year}{2000}\natexlab{}.
\newblock \showarticletitle{Efficient Proofs that a Committed Number Lies in an Interval}. In \bibinfo{booktitle}{\emph{International Conference on the Theory and Application of Cryptographic Techniques}}.
\newblock


\bibitem[Bowe et~al\mbox{.}(2019)]%
        {halo}
\bibfield{author}{\bibinfo{person}{Sean Bowe}, \bibinfo{person}{Jack Grigg}, {and} \bibinfo{person}{Daira Hopwood}.} \bibinfo{year}{2019}\natexlab{}.
\newblock \showarticletitle{Halo: Recursive Proof Composition without a Trusted Setup}.
\newblock \bibinfo{journal}{\emph{IACR Cryptol. ePrint Arch.}}  \bibinfo{volume}{2019} (\bibinfo{year}{2019}), \bibinfo{pages}{1021}.
\newblock


\bibitem[Dagher et~al\mbox{.}(2015)]%
        {provisions}
\bibfield{author}{\bibinfo{person}{Gaby~G. Dagher}, \bibinfo{person}{Benedikt B{\"u}nz}, \bibinfo{person}{Joseph Bonneau}, \bibinfo{person}{Jeremy Clark}, {and} \bibinfo{person}{Dan Boneh}.} \bibinfo{year}{2015}\natexlab{}.
\newblock \showarticletitle{Provisions: Privacy-preserving Proofs of Solvency for Bitcoin Exchanges}.
\newblock \bibinfo{journal}{\emph{Proceedings of the 22nd ACM SIGSAC Conference on Computer and Communications Security}} (\bibinfo{year}{2015}).
\newblock


\bibitem[Davis(2017)]%
        {trumpnyt}
\bibfield{author}{\bibinfo{person}{Julie Davis}.} \bibinfo{year}{2017}\natexlab{}.
\newblock \bibinfo{title}{Trump Won’t Release His Tax Returns, a Top Aide Says}.
\newblock \bibinfo{howpublished}{\url{https://www.nytimes.com/2017/01/22/us/politics/donald-trump-tax-returns.html}}.
\newblock
\newblock
\shownote{Accessed: 5-16-2023}.


\bibitem[de~Chaves and Benitti(2023)]%
        {deChaves2023PrivacyBD}
\bibfield{author}{\bibinfo{person}{Shirlei~Aparecida de Chaves} {and} \bibinfo{person}{Fabiane Barreto~Vavassori Benitti}.} \bibinfo{year}{2023}\natexlab{}.
\newblock \showarticletitle{Privacy by Design in Software Engineering: An update of a Systematic Mapping Study}.
\newblock \bibinfo{journal}{\emph{Proceedings of the 38th ACM/SIGAPP Symposium on Applied Computing}} (\bibinfo{year}{2023}).
\newblock


\bibitem[{e-Estonia}({[n.\,d.]})]%
        {eEstoniaNotAPaper}
\bibfield{author}{\bibinfo{person}{{e-Estonia}}.} \bibinfo{year}{[n.\,d.]}\natexlab{}.
\newblock \bibinfo{title}{e-Tax}.
\newblock \bibinfo{howpublished}{\url{https://e-estonia.com/solutions/ease_of_doing_business/e-tax/}}.
\newblock
\newblock
\shownote{Accessed: 3-16-2023}.


\bibitem[Goldwasser et~al\mbox{.}(1985)]%
        {original}
\bibfield{author}{\bibinfo{person}{Shafi Goldwasser}, \bibinfo{person}{Silvio Micali}, {and} \bibinfo{person}{Charles Rackoff}.} \bibinfo{year}{1985}\natexlab{}.
\newblock \showarticletitle{The knowledge complexity of interactive proof-systems}. In \bibinfo{booktitle}{\emph{Symposium on the Theory of Computing}}.
\newblock


\bibitem[Groth(2010)]%
        {zksnark}
\bibfield{author}{\bibinfo{person}{Jens Groth}.} \bibinfo{year}{2010}\natexlab{}.
\newblock \showarticletitle{Short Pairing-Based Non-interactive Zero-Knowledge Arguments}. In \bibinfo{booktitle}{\emph{International Conference on the Theory and Application of Cryptology and Information Security}}.
\newblock


\bibitem[Groth(2016)]%
        {groth16}
\bibfield{author}{\bibinfo{person}{Jens Groth}.} \bibinfo{year}{2016}\natexlab{}.
\newblock \showarticletitle{On the Size of Pairing-Based Non-interactive Arguments}.
\newblock \bibinfo{journal}{\emph{IACR Cryptol. ePrint Arch.}}  \bibinfo{volume}{2016} (\bibinfo{year}{2016}), \bibinfo{pages}{260}.
\newblock


\bibitem[Gupta(2022)]%
        {gupta2022zk}
\bibfield{author}{\bibinfo{person}{Aayush Gupta}.} \bibinfo{year}{2022}\natexlab{}.
\newblock \bibinfo{booktitle}{\emph{ZK Email}}.
\newblock
\urldef\tempurl%
\url{https://blog.aayushg.com/posts/zkemail}
\showURL{%
\tempurl}


\bibitem[iden3(2023)]%
        {snarkjsgithub}
\bibfield{author}{\bibinfo{person}{iden3}.} \bibinfo{year}{2023}\natexlab{}.
\newblock \bibinfo{title}{zkSNARK implementation in JavaScript \& WASM}.
\newblock \bibinfo{howpublished}{\url{https://github.com/iden3/snarkjs}}.
\newblock
\newblock
\shownote{Accessed: 3-16-2023}.


\bibitem[Intuit({[n.\,d.]})]%
        {intuitAPI}
\bibfield{author}{\bibinfo{person}{Intuit}.} \bibinfo{year}{[n.\,d.]}\natexlab{}.
\newblock \bibinfo{title}{Intuit Developer API Reference}.
\newblock \bibinfo{howpublished}{\url{https://developer.intuit.com/app/developer/qbo/docs/api/accounting/all-entities/taxservice}}.
\newblock
\newblock
\shownote{Accessed: 3-16-2023}.


\bibitem[IRS(2019)]%
        {IRSmodernizationPlan}
\bibfield{author}{\bibinfo{person}{IRS}.} \bibinfo{year}{2019}\natexlab{}.
\newblock \bibinfo{title}{IRS Integrated Modernization Business Plan}.
\newblock \bibinfo{howpublished}{\url{https://www.irs.gov/pub/irs-pdf/p5336.pdf}}.
\newblock
\newblock
\shownote{Accessed: 3-16-2023}.


\bibitem[Jannes et~al\mbox{.}(2023)]%
        {Jannes2023DEDACSDA}
\bibfield{author}{\bibinfo{person}{Kristof Jannes}, \bibinfo{person}{Vincent Reniers}, \bibinfo{person}{Wouter Lenaerts}, \bibinfo{person}{Bert Lagaisse}, {and} \bibinfo{person}{Wouter Joosen}.} \bibinfo{year}{2023}\natexlab{}.
\newblock \showarticletitle{DEDACS: Decentralized and dynamic access control for smart contracts in a policy-based manner}.
\newblock \bibinfo{journal}{\emph{Proceedings of the 38th ACM/SIGAPP Symposium on Applied Computing}} (\bibinfo{year}{2023}).
\newblock


\bibitem[Jie(2019)]%
        {announcingpowerstau}
\bibfield{author}{\bibinfo{person}{Koh~Wei Jie}.} \bibinfo{year}{2019}\natexlab{}.
\newblock \bibinfo{title}{Announcing the Perpetual Powers of Tau Ceremony to benefit all zk-SNARK projects}.
\newblock \bibinfo{howpublished}{\url{https://medium.com/coinmonks/announcing-the-perpetual-powers-of-tau-ceremony-to-benefit-all-zk-snark-projects-c3da86af8377}}.
\newblock
\newblock
\shownote{Accessed: 5-16-2023}.


\bibitem[Kjellstadli et~al\mbox{.}(2019)]%
        {weirdnorwaytaxpeople}
\bibfield{author}{\bibinfo{person}{Espen~Hammer Kjellstadli}, \bibinfo{person}{Mariusz Nowostawski}, \bibinfo{person}{Abylay Satybaldy}, {and} \bibinfo{person}{Nader Aeinehchi}.} \bibinfo{year}{2019}\natexlab{}.
\newblock \showarticletitle{Privacy-preserving tax-case processing}. In \bibinfo{booktitle}{\emph{2019 17th International Conference on Privacy, Security and Trust}}.
\newblock


\bibitem[Ko et~al\mbox{.}(2021)]%
        {ko2021efficient}
\bibfield{author}{\bibinfo{person}{Hankyung Ko}, \bibinfo{person}{Ingeun Lee}, \bibinfo{person}{Seunghwa Lee}, \bibinfo{person}{Jihye Kim}, {and} \bibinfo{person}{Hyunok Oh}.} \bibinfo{year}{2021}\natexlab{}.
\newblock \showarticletitle{Efficient Verifiable Image Redacting based on zk-SNARKs}. In \bibinfo{booktitle}{\emph{Proceedings of the 2021 ACM Asia Conference on Computer and Communications Security}}. \bibinfo{pages}{213--226}.
\newblock


\bibitem[Narula et~al\mbox{.}(2018)]%
        {zkledger}
\bibfield{author}{\bibinfo{person}{Neha Narula}, \bibinfo{person}{Willy Vasquez}, {and} \bibinfo{person}{Madars Virza}.} \bibinfo{year}{2018}\natexlab{}.
\newblock \showarticletitle{zkLedger: Privacy-Preserving Auditing for Distributed Ledgers}. In \bibinfo{booktitle}{\emph{IACR Cryptology ePrint Archive}}.
\newblock


\bibitem[Nichols and Rothstein(2015)]%
        {nichols2015earned}
\bibfield{author}{\bibinfo{person}{Austin Nichols} {and} \bibinfo{person}{Jesse Rothstein}.} \bibinfo{year}{2015}\natexlab{}.
\newblock \showarticletitle{The earned income tax credit}.
\newblock In \bibinfo{booktitle}{\emph{Economics of Means-Tested Transfer Programs in the United States, Volume 1}}. \bibinfo{publisher}{University of Chicago Press}, \bibinfo{pages}{137--218}.
\newblock


\bibitem[Niu et~al\mbox{.}(2022a)]%
        {etaxing}
\bibfield{author}{\bibinfo{person}{Hui-Chong Niu}, \bibinfo{person}{Ting Li}, {and} \bibinfo{person}{Xiugang Gong}.} \bibinfo{year}{2022}\natexlab{a}.
\newblock \showarticletitle{A blockchain-based certifiable anonymous E-taxing protocol}.
\newblock \bibinfo{journal}{\emph{PLoS ONE}}  \bibinfo{volume}{17} (\bibinfo{year}{2022}).
\newblock


\bibitem[Niu et~al\mbox{.}(2022b)]%
        {Niu2022ABC}
\bibfield{author}{\bibinfo{person}{Hui-Chong Niu}, \bibinfo{person}{Ting Li}, {and} \bibinfo{person}{Xiugang Gong}.} \bibinfo{year}{2022}\natexlab{b}.
\newblock \showarticletitle{A blockchain-based certifiable anonymous E-taxing protocol}.
\newblock \bibinfo{journal}{\emph{PLoS ONE}}  \bibinfo{volume}{17} (\bibinfo{year}{2022}).
\newblock


\bibitem[Pertsev et~al\mbox{.}({[n.\,d.]})]%
        {TCwhitepaper}
\bibfield{author}{\bibinfo{person}{Alexey Pertsev}, \bibinfo{person}{Roman Semenov}, {and} \bibinfo{person}{Roman Storm}.} \bibinfo{year}{[n.\,d.]}\natexlab{}.
\newblock \bibinfo{title}{Tornado Cash Privacy Solution - Version 1.4}.
\newblock
\newblock
\newblock
\shownote{Accessed: 3-16-2023}.


\bibitem[Premchand and Choudhry(2018)]%
        {openbanking}
\bibfield{author}{\bibinfo{person}{Anshu Premchand} {and} \bibinfo{person}{Anurag Choudhry}.} \bibinfo{year}{2018}\natexlab{}.
\newblock \showarticletitle{Open Banking \& APIs for Transformation in Banking}.
\newblock \bibinfo{journal}{\emph{2018 International Conference on Communication, Computing and Internet of Things (IC3IoT)}} (\bibinfo{year}{2018}), \bibinfo{pages}{25--29}.
\newblock


\bibitem[Rivera-Zamarripa et~al\mbox{.}(2019)]%
        {mexicoPKI}
\bibfield{author}{\bibinfo{person}{Luis Rivera-Zamarripa}, \bibinfo{person}{Lil~M. Rodriguez}, \bibinfo{person}{Miguel Angel~L{\'e}on Ch{\'a}vez}, \bibinfo{person}{Nareli~Cruz Cort{\'e}s}, {and} \bibinfo{person}{Francisco Rodr{\'i}guez-Henr{\'i}quez}.} \bibinfo{year}{2019}\natexlab{}.
\newblock \showarticletitle{Security Analysis of the Mexican Fiscal Digital Certificate System}.
\newblock \bibinfo{journal}{\emph{Computaci{\'o}n y Sistemas}} (\bibinfo{year}{2019}).
\newblock


\bibitem[Rosenbaum(2013)]%
        {rosenbaum2013relationship}
\bibfield{author}{\bibinfo{person}{Dorothy Rosenbaum}.} \bibinfo{year}{2013}\natexlab{}.
\newblock \showarticletitle{The relationship between SNAP and work among low-income households}.
\newblock \bibinfo{journal}{\emph{Center on Budget and Policy Priorities}} (\bibinfo{year}{2013}).
\newblock


\bibitem[Sanders(2020)]%
        {sanders2020efficient}
\bibfield{author}{\bibinfo{person}{Olivier Sanders}.} \bibinfo{year}{2020}\natexlab{}.
\newblock \showarticletitle{Efficient redactable signature and application to anonymous credentials}. In \bibinfo{booktitle}{\emph{Public-Key Cryptography--PKC 2020: 23rd IACR International Conference on Practice and Theory of Public-Key Cryptography, Edinburgh, UK, May 4--7, 2020, Proceedings, Part II}}. Springer, \bibinfo{pages}{628--656}.
\newblock


\bibitem[Steinfeld et~al\mbox{.}(2002)]%
        {steinfeld2002content}
\bibfield{author}{\bibinfo{person}{Ron Steinfeld}, \bibinfo{person}{Laurence Bull}, {and} \bibinfo{person}{Yuliang Zheng}.} \bibinfo{year}{2002}\natexlab{}.
\newblock \showarticletitle{Content extraction signatures}. In \bibinfo{booktitle}{\emph{Information Security and Cryptology—ICISC 2001: 4th International Conference Seoul, Korea, December 6--7, 2001 Proceedings 4}}. Springer, \bibinfo{pages}{285--304}.
\newblock


\end{thebibliography}

%%
% If your work has an appendix, this is the place to put it.

\appendix

\clearpage
\newpage

\section{Prototype Interfaces}

\begin{figure}[h]
    \includegraphics[width=\linewidth]{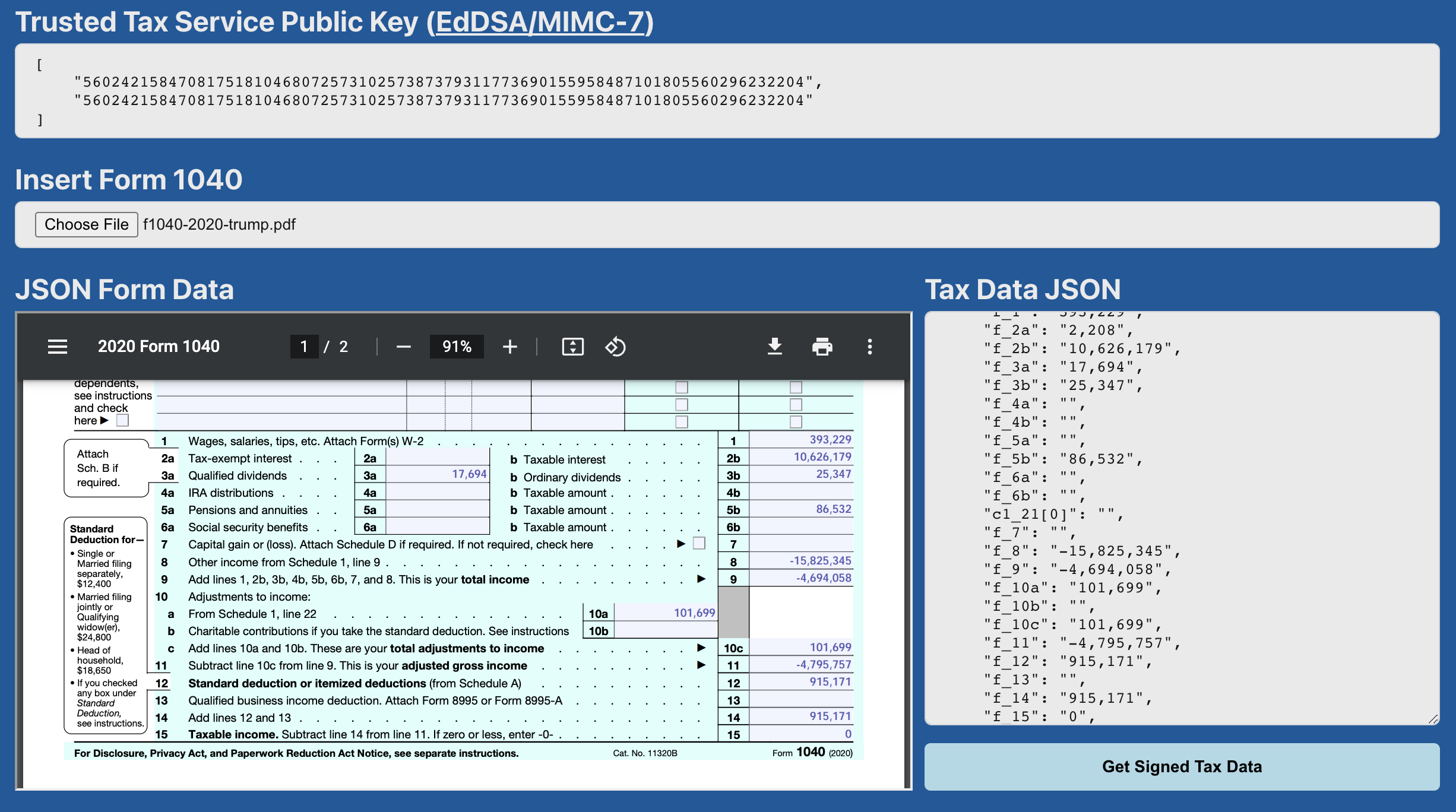}
    \caption{Prototype mock TTS interface. Users can insert Form 1040 and the service returns the JSON representation of the data signed with the TTS key.}
    \label{fig:TTS}
\end{figure}

\begin{figure}[h]
    \includegraphics[width=\linewidth]{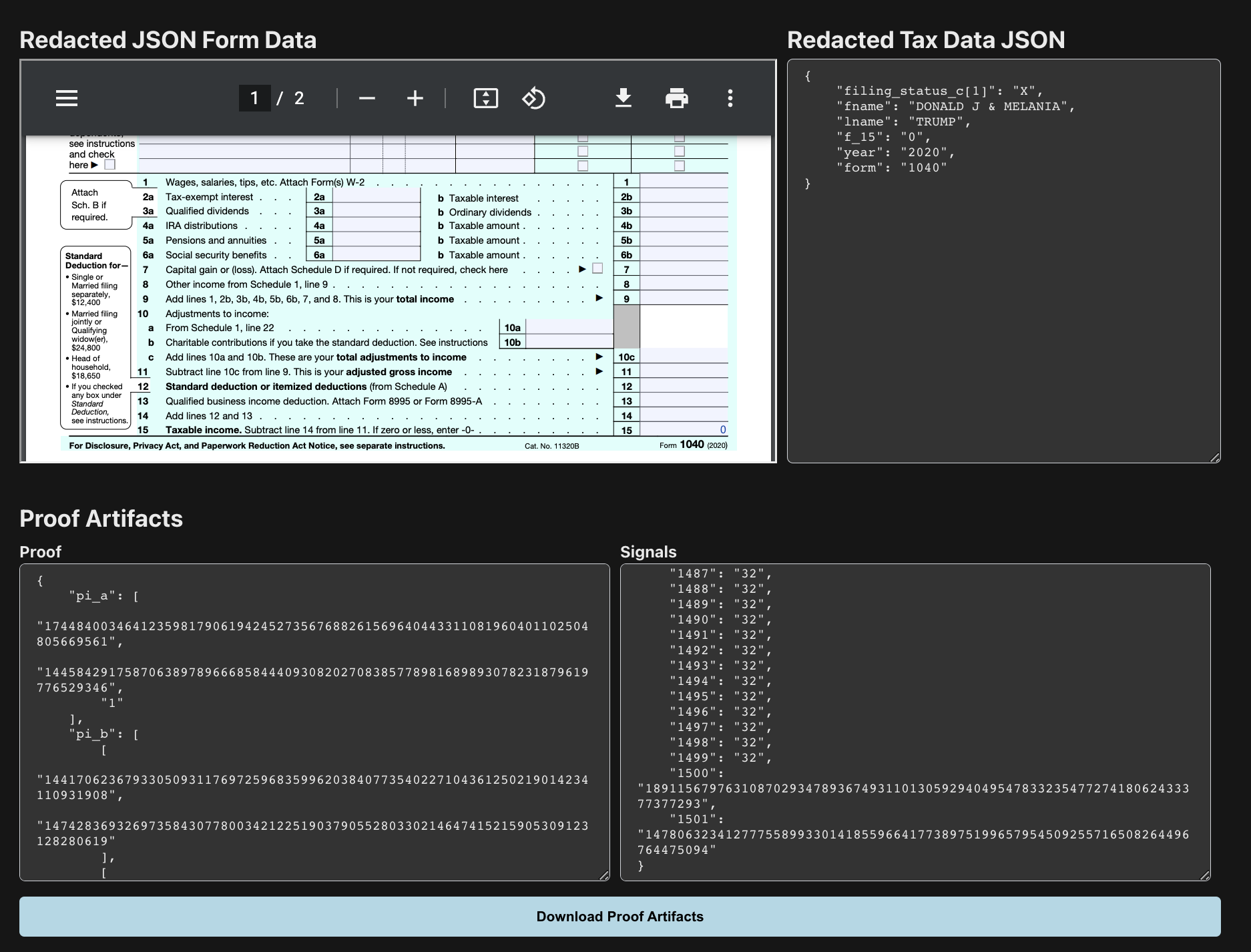}
    \caption{Prototype Redact \& Prove Service interface showing fields relevant to the example use case provided.}
    \label{fig:redact_prove}
\end{figure}

\begin{figure}[h]
    \includegraphics[width=\linewidth]{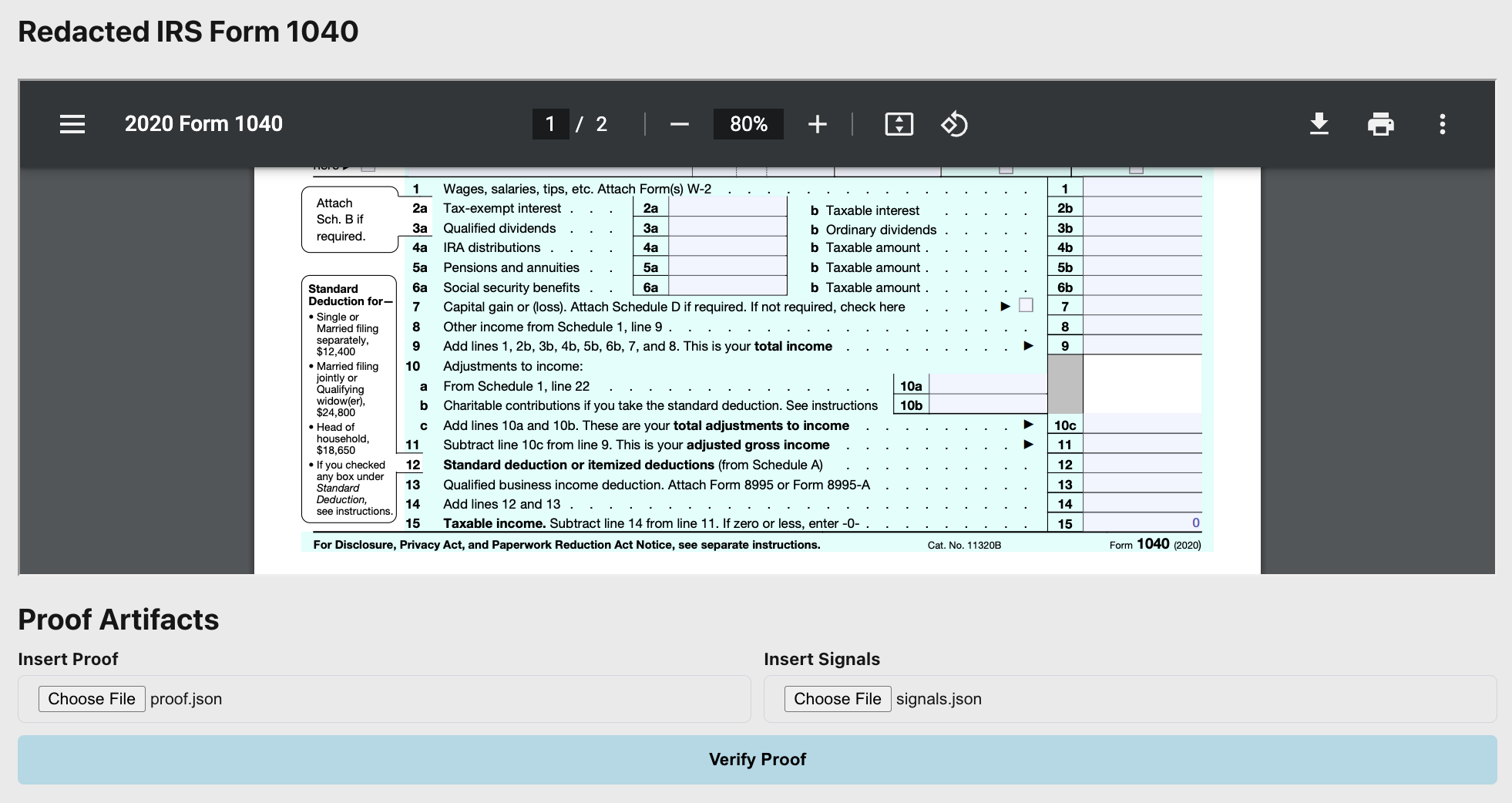}
    \caption{Prototype Verify Service interface showing fields relevant to the example use case provided.}
    \label{fig:verify}
\end{figure}

\begin{figure}[h!]
    \includegraphics[width=\linewidth]{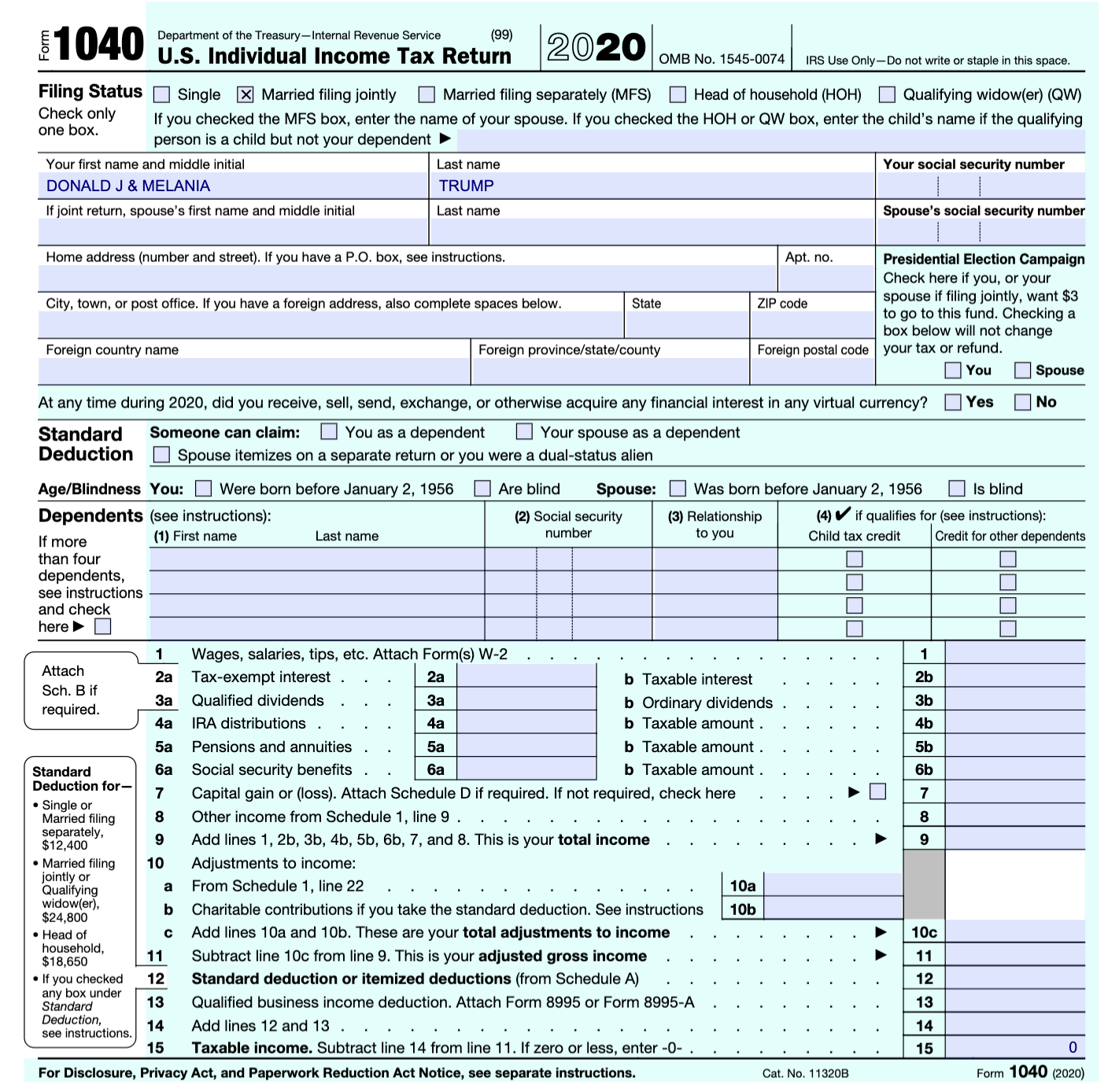}
    \caption{Redacted version of U.S. income tax return Form 1040 for Donald Trump (2020 version), based on publicly released tax returns. All fields are "redacted" (set to blank whitespace) except the requested field 15 and his name (in order to verify his identity for the public release of the redacted version). Note that the form has more pages but the fields relevant for the example are on the first page, so only the first page is shown.}
    \label{fig:trumpf1040_redacted}
\end{figure}

\clearpage

% \begin{figure}
%     \includegraphics[width=\linewidth]{figures/circuit.png}
%     \caption{Code for the circuit written in Circom}
%     \label{fig:circuit}
% \end{figure}

\end{document}